%% file: paper.tex
\documentclass[year=25,pdfa]{fmcad} %

\usepackage[T1]{fontenc}

\usepackage{graphicx} %

\usepackage{adjustbox}
\usepackage{booktabs} %
\usepackage{array} %
\usepackage{paralist} %
\usepackage{proof}
\usepackage{verbatim} %
\usepackage{subfig} %
\usepackage{xcolor}
\usepackage{amsmath}
\usepackage{amssymb}
\usepackage{amsbsy}
\usepackage{amsthm}
\usepackage{algpseudocode}
\usepackage{algorithm}
\usepackage{enumerate}
\usepackage{bussproofs}
\usepackage{centernot}
\usepackage{pifont}
\usepackage{multirow}
\usepackage{balance}
\usepackage{hyperref}

\newcommand{\cmark}{\ding{51}}%
\newcommand{\xmark}{\ding{55}}%
\newcommand{\sygusbench}{I}
\newcommand{\knapsackbench}{II}
\newcommand{\maxbench}{III}
\newcommand{\lowernsbench}{IV}
\newcommand{\lowersbench}{V}
\newcommand{\betweentbench}{VI}
\newcommand{\betweenpbench}{VII}
\newcommand{\equationtbench}{VIII}
\newcommand{\equationpbench}{IX}
\newcommand{\uftbench}{X}
\newcommand{\ufpbench}{XI}

\newcommand\casecmd{\mathtt{case}}
\newcommand\defaultcmd{\mathtt{default}}

\newcommand{\Model}{\mathcal{M}}
\newtheoremstyle{mytheorem}{1pt}{1pt}{\itshape}{}{\bfseries}{.}{.5em}{}
\theoremstyle{mytheorem}
\newcounter{theorem}
{\bfseries}{\itshape}
\newtheorem{Proposition}[theorem]{Proposition}{\bfseries}{\itshape}
{\bfseries}{\itshape}
\newtheorem{Definition}[theorem]{Definition}{\bfseries}{\itshape}
{\bfseries}{\itshape}
{\bfseries}{\itshape}
\newtheorem{Remark}[theorem]{Remark}{\bfseries}{\itshape}
\newtheoremstyle{myexample}{1pt}{1pt}{}{}{\itshape}{.}{.5em}{}
\theoremstyle{myexample}
\newtheorem{example}{Example}
\newtheoremstyle{myproof}{1pt}{1pt}{}{}{\itshape}{.}{.5em}{}
\theoremstyle{myproof}
\newtheorem*{myproof}{Proof}
\makeatletter
\renewcommand{\@upn}{}
\makeatother

\newcommand{\EMBP}{\mbox{$\exists$MBP}}
\newcommand{\EMBPR}{\mbox{$\exists$MBPR}}
\newcommand{\UMBP}{\mbox{$\forall$MBP}}

\newcommand{\realizer}{r}

\newcommand{\Realizer}{R}

\renewcommand{\vec}[1]{\boldsymbol{#1}}
\newcommand{\Uncomputable}{\vec{u}}
\newcommand{\Computable}{\vec{c}}

\newcommand{\Literals}{\mathcal{L}}
\newcommand{\Equalities}{\mathsf{E}}
\newcommand{\Core}{\mathcal{C}}

\newcommand{\Partition}{\mathcal{P}}
\newcommand{\UProject}{\pi^\forall}
\newcommand{\EProject}{\pi^\exists}

\newcommand{\Sunday}{\mathit{Sun}}
\newcommand{\Monday}{\mathit{Mon}}
\newcommand{\Terms}{\mathcal{T}}

\newcommand{\lo}[1]{\mathrm{lo}(#1)}
\newcommand{\hi}[1]{\mathrm{hi}(#1)}
\newcommand{\val}[1]{\mathrm{val}(#1)}

\newcommand{\reprC}{\mathsf{rep}_{\Computable}}

\newcommand{\justify}[2]{\mathcal{J}(#1, #2)}
\newcommand{\ComputableTerms}{T_{\Computable}}

\newcommand{\SigmaTerms}{T_{\Sigma}}
\newcommand{\ComputableLiterals}{\Literals_{\Computable}}
\newcommand{\UncomputableLiterals}{\Literals_{\Uncomputable}}
\newcommand{\Tableau}{T}
\newcommand{\Bounds}{B}
\newcommand{\ProjectionClauses}{\mathcal{O}}
\newcommand{\CorrectionSets}{\mathcal{S}}
\newcommand{\NNF}{\mathit{NNF}}

\begin{document}
\title{Synthesiz3 This: an SMT-Based Approach for Synthesis with Uncomputable Symbols}
\author{\IEEEauthorblockN{Petra Hozzov\'a\IEEEauthorrefmark{1}\orcid{0000-0003-0845-5811}
and
Nikolaj Bj\o{}rner\IEEEauthorrefmark{2}\orcid{0000-0002-1695-2810}
}
  \IEEEauthorblockA{\IEEEauthorrefmark{1}Czech Technical University, Prague, Czech Republic \\
\IEEEauthorrefmark{2}Microsoft, Redmond, USA
  }
}

\maketitle

\begin{abstract}
  Program synthesis is the task of automatically constructing a program conforming to a given specification.
  In this paper we focus on synthesis of single-invocation recursion-free functions conforming to a
  specification given as a logical formula in the presence of uncomputable symbols (i.e., symbols used in the specification but not allowed in the resulting function).
  We approach the problem via SMT-solving methods: we present a quantifier elimination
  algorithm using model-based projections for both total and partial function synthesis,
  working with theories of uninterpreted functions and linear arithmetic and their combination.
  For this purpose we also extend model-based projection to produce witnesses for these theories.
  Further, we present procedures tailored for the case of uniquely determined solutions.
  We implemented a prototype of the algorithms using the SMT-solver Z3, demonstrating their practical efficiency compared to the state of the art. %
\end{abstract}

\input{introduction}

\input{preliminaries}

\input{overview}

\input{eufprojection}

\input{unique}
\input{implementation}

\input{related}

\section{Conclusions}\label{sec:conclusions}
This paper formally defines and studies the problem of synthesis of total and partial functions in the presence of uncomputable symbols.
To solve these problems, we introduced a modular algorithm for synthesis, relying on quantifier projections.
We summarized requirements on the projections and presented %
algorithms computing existential projection with witnesses for EUFLIA, and universal projection internally relying on existential projection.
Further, we described a method to circumvent quantifier solving for uniquely defined functions in \mbox{EUFLRA}.
Finally, we implemented a prototype demonstrating that the proposed methods work in practice.

Our results prompt future work in multiple directions, such as
extending synthesis to functions with different inputs, or
extending the specification with quantified background axioms to support reasoning with user-defined theories.
Moreover, developing projections for other theories would allow synthesis in those theories by plugging them into our algorithm.
Last but not least, our unique solution approach could be extended to infer all implied equalities.

\medskip

\noindent {\bf Acknowledgments.} We thank Kuldeep Meel for suggesting to examine partial function synthesis with unique realizers.
Diego Olivier Fernandez Pons inspired us to explore synthesis with Z3.
We also thank Mikol\'a\v{s} Janota for discussions of examples.
Petra Hozzov\'a was funded by the European Union under the project ROBOPROX (reg. no. CZ.02.01.01/00/22\_008/0004590).
\balance

\bibliographystyle{plain}
\bibliography{refs}

\end{document}

%% file: introduction.tex
\section{Introduction}

Program synthesis is the task of finding a function satisfying the given specification.
In this paper we focus on synthesis of recursion-free functions and require the solution to be guaranteed to be correct.
To this end we lean on automated reasoning methods and do not consider machine learning.
As synthesis subsumes verification, it is an inherently hard problem.
Yet, recent developments in this area showed that fragments of the problem are tractable~\cite{GulwaniEtAl2011,TiwariEtAl2015,DBLP:journals/fmsd/ReynoldsKTBD19,PetraCADE23}.

There are many flavors of synthesis specifications.
We consider specifications expressed as logical formulas capturing relational constraints on symbolically represented inputs/outputs, recently addressed by an SMT-based approach~\cite{DBLP:journals/fmsd/ReynoldsKTBD19} and deductive synthesis in saturation~\cite{PetraCADE23}.
Such specifications can use additional syntactic constraints: %
in~\cite{DBLP:journals/fmsd/ReynoldsKTBD19} (and in the SyGuS paradigm in general~\cite{DBLP:series/natosec/AlurBDF0JKMMRSSSSTU15}), the resulting function must syntactically conform to the provided grammar,
while~\cite{PetraCADE23} introduces so-called \emph{uncomputable} symbols
which are not allowed to occur in the solution. %
However, while~\cite{PetraCADE23} uses uncomputable symbols, it does not formally characterize them.
That motivated this work: we explore the semantics of specifications with uncomputable symbols,
and we present an algorithm solving such specifications modulo (combinations of) theories.

Uncomputable symbols are a powerful tool allowing to make the specification more expressive, e.g. by
(i) requiring the solution to be maximal, %
(ii) assuming the necessary condition for the existence of a total solution, %
or (iii) forcing the solution to discover properties of functions or predicates used in the specification, but not allowed in the output.
While~\cite{PetraCADE23} introduces specialized inference rules to handle the uncomputable symbols, we integrate them directly into the specification, expressed as a logical formula, by using second-order quantification.
The challenges we then solve are how to \emph{project away} these symbols from the formulas,
and how to \emph{find witnesses} capturing the solutions.
To solve these challenges, we take an SMT-based route:
we present an algorithm based on \emph{quantifier elimination games}~\cite{DBLP:conf/lpar/BjornerJ15,DBLP:conf/ijcai/FarzanK16},
using and extending \emph{model-based projections}~\cite{DBLP:conf/cav/KomuravelliGC14,EasyChair:10000}.

\paragraph*{Contributions}
In this work we %
synthesize recursion-free functions for first-order specifications using uncomputables and theories.
Our main contributions are:
\begin{enumerate}
  \item An algorithm for function synthesis in the presence of \emph{uncomputable symbols} (Section~\ref{sec:overview}).
    In contrast to prior work%
    ~\cite{PetraCADE23}, we use quantifiers to handle the scope of uncomputables. %
    We present tailored model-based projection procedures (Section~\ref{sec:mbp}) used for quantifier elimination.
\item Our approach works directly for \emph{partial function synthesis} (see Section~\ref{sec:partial-intro}). That is, for specifications that are not realizable on all inputs, our approach identifies necessary and sufficient \emph{pre-conditions} when solutions exist.
\item %
  We introduce directed inferences exploiting the case when specifications have \emph{unique solutions} (Section~\ref{sec:unique-functions}).
\item We implemented a prototype of our method using the SMT-solver Z3 and compared it to the state of the art (Section~\ref{sec:implementation}).
\end{enumerate}

%% file: preliminaries.tex
\section{Preliminaries}\label{sec:preliminaries}
We assume multi-sorted first-order and second-order logics and use standard definitions of their syntax and semantics. %
When discussing validity, we consider all formulas implicitly universally quantified.
We work with theories as defined in SMT-LIB~\cite{SMTLIB}, in particular with the theories of equality and uninterpreted functions (EUF), and of linear integer and real arithmetic (LIA and LRA).

By $E[t]$ we signify that the term or formula $E$ contains the term $t$,
and by $E[t/y]$ we denote the term or formula we obtain from $E$ by replacing every occurrence of the variable $y$ by the term $t$.
By $\simeq$ we denote the equality predicate, by $:=$ assignment,
by $\vec{t}$ the tuple of terms (or variables/symbols) $t_1,\dots,t_n$,
and by $\SigmaTerms$ the set of terms over the set of symbols $\Sigma$.
We recall known notions we use in the paper:
negation normal form (NNF) of a formula $F$, denoted $\NNF(F)$, is any formula equivalent to $F$, consisting only of literals, conjunctions, and disjunctions;
disjunction normal form (DNF) is an NNF which is a disjunction of conjunctions of literals.
We use the abbreviations sat/unsat for satisfiable/unsatisfiable.
We abuse the notation and by $\top, \bot$ we denote the logical constants for true and false, as well as their values, and in our algorithms we also denote the undefined value by $\bot$.
We reduce predicates into functions by replacing each atom $p(\vec{t})$ by $f_p(\vec{t})\simeq\top$, where $f_p$ is a fresh function symbol with the domain $\{\bot,\top\}$.
We write $\bar{\ell}$ for the complement of the literal $\ell$.
When $C$ is a conjunction of literals $\bigwedge_{\ell\in\Literals}\ell$, we interchangeably refer to it as a set of its literals $\Literals$.
Given a set of literals $\Literals$ and a formula $F$, where $\Literals\land F$ is unsat, we write $\mathit{Core}(\Literals, F)$ for %
$\Core\subseteq\Literals$ such that $\Core\land F$ is unsat.

We use ``program'' and ``function'' in the context of objects we synthesize as synonyms.
By $\epsilon$ we denote the empty list and we use $|$ to denote list construction.
We write $C\rightarrow r$ for the pair of a condition $C$ and a term $r$ used as one case of a program,
and we represent the (possibly partial) program in the synthesis process as a list $C_1\rightarrow r_1~|~\dots~|~ C_n\rightarrow r_n$.
We construct the whole program from the list using the case constructor,
\[\casecmd\ C_1:r_1;\ \dots\ \casecmd\ C_{n-1}:r_{n-1};\ \defaultcmd: r_n,\]
which evaluates to the value $r_i$ for $1\leq i< n$ if $C_i\land \bigwedge_{j=1}^{i-1}\neg C_j$ is true,
and to the value $r_n$ otherwise.
Since a list of pairs of conditions and terms uniquely defines a program, we sometimes abuse the notation and use the list in place of its corresponding program.
Finally, we use $\_$ to denote \emph{any} value in the function -- i.e., the case when any value of the correct sort satisfies the specification, and we use $\_$ when there exists at least one computable term of that sort.

%% file: overview.tex
\section{Synthesis of Total and Partial Functions}\label{sec:overview}

\subsection{Specifications and Realizers}

We focus on the class of synthesis specifications given as closed second-order logical formulas of the form
\begin{equation}\label{eq:spec-f}
  \exists f.\forall\Computable,\Uncomputable.\ \varphi[f(\Computable), \Computable,\Uncomputable],
\end{equation}
where $f$ is the recursion-free function we want to synthesize, $\Computable$ are all symbols the function can depend on, and $\Uncomputable$ are the symbols $f$ should be independent of.
Following the terminology of~\cite{PetraCADE23}, we call $\Computable$ \emph{computable} and $\Uncomputable$ \emph{uncomputable} symbols.
The computable symbols include all symbols which are allowed in the solution for $f$:
$f$'s inputs, %
as well as constants, %
functions, and predicates. %
We call the terms and formulas over $\Computable$ \emph{computable terms/formulas}, the terms/formulas containing symbols from $\Uncomputable$ \emph{uncomputable terms/formulas}, and the uncomputable symbols themselves also \emph{uncomputables}.

Note that~\eqref{eq:spec-f} specifies a \emph{single single-invocation function}: $f$ always occurs in $\varphi$ with the same arguments $\Computable$.
Thus, specification~\eqref{eq:spec-f} can be equivalently expressed in ``unskolemized'' (yet still possibly second-order) form as
\begin{equation}\label{eq:spec-y}
  \forall\Computable.\exists y.\forall\Uncomputable.\ \varphi[y, \Computable,\Uncomputable],
\end{equation}
where the task is to find a witness for the first-order variable $y$.
We can also allow $y$ to be a tuple of variables, effectively specifying multiple functions, all depending on the same computable symbols/inputs.

In the rest of the paper, we use specifications like~\eqref{eq:spec-y}, %
formalized as follows:
\begin{Definition}[Total Function Synthesis]
Let $\Sigma$ be a signature,
let $\Uncomputable\subseteq\Sigma$ be the set of uncomputable symbols, %
let $y\in\Sigma\setminus\Uncomputable$,
and let $\Computable = \Sigma\setminus(\{y\}\cup\Uncomputable)$.
A \emph{synthesis problem} is a tuple
\begin{equation*} \label{eq:synth-spec}
  \langle \Phi[y], \Uncomputable, y \rangle,
\end{equation*}
where $\Phi[y] = \varphi[y, \Computable, \Uncomputable]$ is a quantifier-free formula called a \emph{synthesis specification}, %
containing the variable $y$, which represents the function to be synthesized.
A solution for the problem is %
a term $\Realizer\in\ComputableTerms$, such that $\Phi[\Realizer/y]$ holds, called a \emph{realizer} for $y$.
As $\Realizer$ defines a total function, we call this problem \emph{total function synthesis}.
\end{Definition}

We consider theory-defined functions computable.
In the following we always use $y$ to represent the function to be synthesized, and
sometimes write $\Phi$ for $\Phi[y]$, and when $t$ does not contain $y$, we write $\Phi[t]$ for $\Phi[t/y]$.

\begin{example}[2-knapsack]\label{ex:running1}
  We encode the well-known problem:
  Given 2 items of weights $w_1, w_2$ and a capacity $c$, which of the items should we choose so that the total weight of the chosen items is maximal but at most $c$?
  First, we define a formula specifying that $y_1, y_2$ form a valid pick from the two items, i.e., that the value of $y_i$ is either 0 or $w_i$ (for $i = 1,2$), and that $y_1+y_2$ is at most $c$:
      \begin{align*}%
        \varphi[y_1, y_2] &:= (y_1 \simeq  0\lor y_1\simeq w_1) \land (y_2 \simeq  0\lor y_2\simeq w_2) \\ & \qquad   \land y_1+y_2 \leq c
      \end{align*}
      Then the following synthesis problem specifies, assuming that $c, w_1, w_2$ are non-negative, that $\vec{y}$ is a valid pick from the two items, such that all valid picks $\Uncomputable$ have lesser or equal combined weight than $\vec{y}$:
      \begin{align}\label{eq:2-knapsack-spec}
        \begin{split}
          &\langle \big(c\!\geq\! 0\land w_1\!\geq\! 0\land w_2\!\geq\! 0\big)\\
          &\quad \!\implies\!%
            \big(\varphi[\vec{y}] \land (\varphi[\vec{u}]\!\implies\!u_1\!+\!u_2\!\leq\!y_1\!+\!y_2)\big),\ \{\vec{u}\},\ \vec{y} \rangle
        \end{split}
      \end{align}
  A realizer for $(y_1, y_2)$ of~\eqref{eq:2-knapsack-spec} is:
\begin{align*}
  &\casecmd\ w_1+w_2\leq c : (w_1, w_2);\\
  &\casecmd\ w_1\leq c \land (w_1\geq w_2\lor w_2 > c) : (w_1, 0);\\
  &\casecmd\ w_2\leq c \land (w_2>w_1\lor w_1 > c) : (0, w_2);\\
  &\defaultcmd: (0, 0)
\end{align*}
\end{example}

\subsection{The Expressivity of the Uncomputable Symbols}
We define the scope of uncomputable symbols directly in the specification, by using explicit (and possibly second-order) quantification.
This is in contrast with~\cite{PetraCADE23}, which introduced uncomputable symbols, but forbade their use in the sought solution as an additional syntactic constraint, thus allowing the specifications to stay within first-order logic.
These two methods of restricting the uncomputable symbols are, however, equivalent in what they allow to express.

The use of uncomputable symbols allows for more expressive specifications, which in turn leads to more interesting synthesis tasks.
As in Example~\ref{ex:running1}, with uncomputables we can require that the solution is optimal
by changing the specification $\varphi[y, \Computable]$
into $\varphi[y, \Computable]\land (\varphi[u, \Computable]\implies \mathit{is\_better}(y, u))$, using a fresh uncomputable symbol $u$ and some suitable definition of the predicate $\mathit{is\_better}$.
We can also use uncomputable symbols to ensure that a specification $\varphi[y, \Computable]$ has a solution by adding an assumption $\varphi[u, \Computable]$, allowing to turn partial function specifications into total ones.
For example, there is no total solution for the specification $\langle 2y \simeq x, \emptyset, y\rangle$ over integers, but there is one for
$\langle 2u\simeq x \implies 2y\simeq x, \{u\}, y\rangle$.\footnote{This trick does not work if $\varphi$ already uses some uncomputable symbols.
However, our synthesis algorithm (Section~\ref{sec:algorithm-setup}) works for partial functions, as defined in Section~\ref{sec:partial-intro}.}

Further, $\Uncomputable$ can include also function and predicate symbols.
With those, synthesis can require discovering properties of uninterpreted functions or predicates:
\begin{example}[Workshop~\cite{Reger2018,VampireWS23}]\label{ex:workshop-spec}
  Consider the synthesis problem $\langle\Phi, \{w\}, y\rangle$, where:
  \begin{align*}\label{eq:workshop-spec}
    \Phi &:= \big(
    (\Monday \!\lor\! \Sunday) \land
    (\Monday \!\implies\! w(\mathit{V})) \land
    (\Sunday \!\implies\! w(\mathit{A}))     
    \big) \\ & \qquad \implies w(y)
  \end{align*}
  The example models a situation at a conference:
  given that it is Monday or Sunday today, and given that the workshop $A$ is taking place on Sunday and the workshop $V$ on Monday, the task is to synthesize a function that tells us which workshop is taking place.
  Additionally, the solution is not allowed to use the workshop predicate $w$ -- if we were able to evaluate $w(\cdot)$ ourselves, we arguably would not need to use any tools to synthesize the solution.
\end{example}

\subsection{Partial Function Synthesis}\label{sec:partial-intro}
We further consider \emph{partial function synthesis}: a variation on the synthesis problem where the specification is not required to be satisfiable on all evaluations of computable symbols.
Thus, the sought function is not required to be total.

\begin{Definition}[Partial Function Synthesis]\label{def:partial-synth}
  Let $\Phi[y]$ be a specification, where $\forall\Computable.\exists y.\forall\Uncomputable.\Phi[y]$ is not necessarily valid.
  Then $\langle\Phi[y],\Uncomputable,{y}\rangle$ is a \emph{partial (function) synthesis problem}.
  A \emph{solution} for the problem  %
  is a pair $\langle C, \Realizer \rangle$,
where $C$ is a quantifier-free formula and $\Realizer$ is a realizer for $y$, both over $\Computable$, such that:
\begin{eqnarray}
  C & \implies & \Phi[\Realizer] \label{eq:sufficient-condition} \\
  (\exists y. \forall \Uncomputable. \Phi) & \implies & C \label{eq:maximal-condition}
\end{eqnarray}
\end{Definition}

The first condition ensures that $\Phi[\Realizer]$ holds under $C$,
and the second condition ensures that $\langle C, \Realizer\rangle$ is a \emph{maximal solution for partial function synthesis}.

While solutions $\langle C, \Realizer\rangle$ are recursively enumerable for first-order specifications,
we note that they are in the undecidable $\forall\exists\forall$-fragment~\cite{Kahr1962}.
The process of finding $C$ that satisfies condition \eqref{eq:maximal-condition}
relates to synthesizing uniform quantifier-free interpolants~\cite{DBLP:conf/cade/YorshM05}.
\begin{example}\label{ex:qy_or_aby}
  Let $\Phi := q(y)\lor a\simeq b\simeq y$ with $\Uncomputable=\{q\}, \Computable=\{a, b\}$.
  Then the weakest quantifier-free formula that implies $\forall\Uncomputable.\Phi$ is $a \simeq b \simeq y$,
  producing the partial solution $\langle a \simeq b, a \rangle$.
  On the other hand, for $\Phi := p(u) \lor a \simeq b\simeq y$, with $\Uncomputable=\{u\}, \Computable=\{p,a, b\}$,
  the strongest $C\in\ComputableTerms$ implied by $\forall\Uncomputable.\Phi$ is $\top$, so while $\langle a \simeq b, a\rangle$
  satisfies~\eqref{eq:sufficient-condition}, it does not satisfy~\eqref{eq:maximal-condition}.
\end{example}
Finite quantifier-free uniform interpolants do not always exist for combinations of theories: 
\begin{example}\label{ex:no-finite-qf-interpolant}
  Let $\Phi := a \simeq y \land \Psi$, where $\Psi$ does not
  have a uniform quantifier-free interpolant.  For example,
  let $\Psi$ be $\neg (p(u) \land a < u < b)$ for integers
  $u, a, b$, where $\Uncomputable = \{u\}$.  Then there
  is no finite maximal solution to the synthesis problem, only an
  infinite one: $\langle \bigvee_{k>1}(a + k \simeq b \land \bigwedge_{0 < i
    < k} \neg p(i + a)),\ a \rangle$.
\end{example}
Thus, we cannot develop complete algorithms in general.
Yet, for theories admittig uniform interpolation, or even quantifier elimination, computing the strongest $C$
satisfying~\eqref{eq:maximal-condition} is possible.
Our synthesis algorithm, presented next, finds partial solutions to \eqref{eq:sufficient-condition}
and uses \eqref{eq:maximal-condition} as a criterion for termination.

\subsection{A Synthesis Algorithm}
\label{sec:algorithm-setup}

We present an algorithm for partial function synthesis.
It iteratively builds a list of pairs $\Realizer := C_1 \rightarrow r_1 ~|~ \dots ~|~ C_n \rightarrow r_n$, stopping when $\forall\Uncomputable. \Phi\land \bigwedge_{1\leq i \leq n} \lnot C_i$ becomes unsatisfiable (step~2) -- i.e., when the conditions $C_1,\dots, C_n$ cover all possible cases.
Then the solution is $\langle \bigvee_{1\leq i\leq n} C_i, \Realizer\rangle$.
To compute one pair $C_j \rightarrow r_j$, we first find a formula $\UProject[y]$ not containing symbols from $\Uncomputable$, which implies $\forall\Uncomputable.\Phi\land\bigwedge_{1\leq i \leq j-1}\lnot C_i$ (step~3).
This $\UProject[y]$ corresponds to a condition under which the specification holds, yet none of the other conditions found so far apply.
Next we look for a witness $\realizer$ and a formula $\EProject$, both only containing symbols from $\Computable$, such that $\EProject$ implies $\UProject[\realizer]$ (step~4).
If we find such $\realizer$ and $\EProject$, we set $C_j := \EProject, r_j := \realizer$ and add $C_j\rightarrow r_j$ to the list of pairs $\Realizer$ (step~6).
Otherwise we find a formula $\EProject$ which implies $\exists y.\UProject[y]$, and refine the specification by adding $\lnot\EProject$ to it (step~5).
Then we continue with the next iteration: if $\forall\Uncomputable. \Phi\land \bigwedge_{1\leq i \leq j}\lnot C_i$ is satisfiable, we look for $C_{j+1}\rightarrow r_{j+1}$.
Computation of $\UProject$ and $\EProject$ (steps 3 and 4) is addressed in Section~\ref{sec:mbp}.

The algorithm can also be used for total function synthesis by checking that the returned condition $C$ is equivalent to $\top$.

\begin{algorithm}
Given a partial function synthesis task $\langle \Phi, \Uncomputable, y \rangle$, return a solution $\langle C, \Realizer\rangle$.
\begin{enumerate}
  \item Initialize $C := \bot$ and $\Realizer := \epsilon$.
\item If $\forall \Uncomputable \ . \ \Phi \land \neg C$ is unsat then return $\langle C, \Realizer\rangle$.
\item Find a satisfiable $\UProject[y]$ over $\Computable\cup\{y\}$, such that $\UProject[y] \implies \forall \Uncomputable \ . \ \Phi \land \neg C $.
  (See Algorithm~\ref{alg:universal-project-loop} in Section~\ref{sec:universal-projection}.)
\item Find $\langle \EProject, \realizer\rangle$ such that $\EProject$ is satisfiable and over $\Computable$ and either $\EProject \implies \UProject[\realizer]$ with $\realizer\in\ComputableTerms$,
  or $\realizer=\bot$ and there is no realizer over $\Computable$ for $\UProject[y]$ and
  $\EProject \implies \exists y . \UProject[y]$. \\
    (See Algorithm~\ref{alg:embpr-euf} in Section~\ref{sec:existential-projection}.)
\item %
  If $\realizer=\bot$, update $\Phi := \Phi \land \neg \EProject$ and go to step 2.
\item Otherwise, update $C := C \lor \EProject$, $\Realizer := \Realizer \mid (\EProject \rightarrow \realizer)$ and go to step 2.
\end{enumerate}
\caption{Partial Function Synthesis~\label{alg:skolem-synthesis}}
\end{algorithm}

Correctness of Algorithm~\ref{alg:skolem-synthesis} is straightforward:

\begin{Proposition}\label{prop:synthesis-correctness}
  Assume Algorithm~\ref{alg:skolem-synthesis} terminates with $\langle C, \Realizer\rangle$.
  If it never produced $\realizer=\bot$ in step 4, then $\langle C, \Realizer\rangle$ is a maximal solution for the given partial synthesis problem.
  Otherwise $\langle C, \Realizer\rangle$ is a program satisfying condition~\eqref{eq:sufficient-condition}
  but there is no solution to~\eqref{eq:maximal-condition}.
\end{Proposition}

\begin{myproof}
  Each step preserves \eqref{eq:sufficient-condition} $C \implies \Phi[\Realizer]$.
  The termination condition ensures \eqref{eq:maximal-condition} $\exists y. \forall \Uncomputable.\Phi \implies C \vee C^\bot$, where $C^\bot$ is the disjunction of projections for $\realizer = \bot$.
\end{myproof}

\begin{Remark}\label{remark:completeness}
  Algorithm~\ref{alg:skolem-synthesis} is complete for theories that admit quantifier elimination
  for computing $\UProject, \EProject$ in steps 3 and 4.
\end{Remark}

\begin{example}\label{ex:workshop}
  Our algorithm solves the workshop problem from Example~\ref{ex:workshop-spec}.
  In the first iteration, $\forall w.\Phi$ is sat.
  We find $\UProject[y] = \neg\Monday\land\neg\Sunday$ (see Example~\ref{ex:workshop-umbp} for details).
  Then a corresponding $\langle\EProject,\realizer\rangle$ is $\langle \neg\Monday\land\neg\Sunday, \_\rangle$, where $\_$ is the ``any'' realizer.
  We add $\neg\Monday\land\neg\Sunday$ to $C$.
  As $\forall w. \Phi\land\neg C$ is satisfiable, we continue.
  We find $\UProject[y] = A\simeq y\land \Sunday$, and then a corresponding $\langle\EProject,\realizer\rangle = \langle\Sunday, A\rangle$ (see Example~\ref{ex:workshop-embpr}).
  We add $\neg\Sunday$ to $C$.
  The formula $\forall w. \Phi\land\neg C$ is still satisfiable, so we continue.
  In the next iteration, $\UProject = V\simeq y\land\Monday$ and $\langle\EProject, \realizer\rangle = \langle\Monday, V\rangle$.
  Then finally $\forall w.\Phi\land\neg C$ is unsat, and in fact, $C\equiv \top$.
  We thus return
  \[\langle \top, (\neg\Monday\land\neg\Sunday\rightarrow\_)~|~(\Sunday\rightarrow A)~|~(\Monday\rightarrow V)\rangle.\]
  Since the first case in the realizer list uses %
  $\_$, we can choose to return $V$ in that case.
  Further, since its condition $\neg\Monday\land\neg\Sunday$ is the complement of the disjunction of the other two conditions ($\Sunday$ or $\Monday$), we can move it to the end of the list and then unite into a $\defaultcmd$ case with the condition $\Monday$, which also returns $V$, obtaining the final program:
  \[\langle \top,\ \casecmd\ \Sunday: A;\ \defaultcmd : V \rangle\]
\end{example}

\subsection{Unique Realizers}
A useful special case in step 4 of Algorithm~\ref{alg:skolem-synthesis} is when a solution to the variable $y$ is uniquely determined.
For example, for a simplified 1-item version of knapsack, %
given by $c<0\lor w_1<0\lor((y_1=0\lor y_1=w_1)\land y_1\leq c\land((u_1\not=0\land u_1\not=w_1)\lor u_1>c\lor u\leq y))$,
and under the case $c\geq 0,w_1\geq 0, u_1=w_1,u_1 > c$, there is a unique solution to $y$ given by $0$.
In Section~\ref{sec:unique-functions}, we show how to read out unique solutions from a combination of the E-graph (a data structure capturing congruences, see Definitions~\ref{def:cpp} and~\ref{def:egraph}) and
arithmetical constraints. Our approach can be considered a generalization of E-graph saturation, where we
consider also the theory of linear real arithmetic.

%% file: eufprojection.tex
\section{Quantifier Projection}
\label{sec:mbp}
Our synthesis algorithm (Algorithm~\ref{alg:skolem-synthesis}) uses two flavors of
quantifier projection methods, one for existential and one for universal quantifiers.
Prior work~\cite{EasyChair:10000} explores projection operators for EUFLIA, but side-steps some requirements
we have for the resulting formulas and realizers.
We therefore introduce heuristics suitable for our setting.
We describe methods %
for existential projection in Section~\ref{sec:existential-projection},
and Section~\ref{sec:universal-projection} presents an approach for universal projection.
We then plug these methods into Algorithm~\ref{alg:skolem-synthesis}.
For step 3 of Algorithm~\ref{alg:skolem-synthesis}, we use universal projection (Algorithm~\ref{alg:universal-project-loop}),
which takes $\Phi\land \lnot C$ and returns a set of literals $\Literals$, which we assign to $\UProject$.
For step 4 of Algorithm~\ref{alg:skolem-synthesis}, we use existential projection with witness extraction for the appropriate theory (for EUF, see Algorithm~\ref{alg:embpr-euf}), which takes $\UProject$ and an arbitrary model of it, and returns a suitable pair $\langle\EProject, r\rangle$.

\subsection{Existential Projection}
\label{sec:existential-projection}
To compute realizers, we use existential model-based projection.

\begin{Definition}[\EMBP{} and \EMBPR~\cite{DBLP:conf/cav/KomuravelliGC14}]
  Let $\Phi$ be a quantifier-free formula and assume $\Model \models \Phi$.
  Then \emph{existential model-based projection} $\EMBP(\Model, \vec{x}, \Phi)$ is a quantifier-free formula $\EProject$ over symbols not in $\vec{x}$, such that
  $\Model \models \EProject$
  and $\EProject\implies \exists \vec{x}. \Phi$.\\
  Further, when $y$ is a symbol in $\Phi$,
  a \emph{model-based projection with realizer}
  $\EMBPR(\Model, y, \Phi)$ is a pair $\langle\EProject, \realizer\rangle$ such that
  $\EProject$ is quantifier-free and satisfiable,
  $\EProject \implies \Phi[\realizer/y]$, and $y\not\in\EProject,\realizer$.
\end{Definition}

Various algorithms for model-based projection have been integrated into Z3.
We will use \EMBP{} to project uncomputable
symbols in Section~\ref{sec:universal-projection}.
We also introduce an algorithm for computing $\EMBPR(M, y, \Literals)$ for EUF given a set of literals $\Literals$.
Prior works~\cite{EasyChair:10000,Komuravelli2015} develop projection algorithms that apply to combinations with arrays~\cite{ArieArray}, 
integer arithmetic, and algebraic data-types~\cite{DBLP:conf/lpar/BjornerJ15}.
However, they do not cover extracting computable realizers from projections.
For example, \cite{EasyChair:10000} defines projections of first-order functions that can be defined without getting into notions of computable realizers.
On the other hand, we can rely on witness extraction algorithms for LIA~\cite{DBLP:conf/pldi/KuncakMPS10}
for a combination with EUF.

\begin{Definition}[Congruence Preserving Partition]\label{def:cpp}
  Let $\Terms$ be a set of terms closed under sub-terms.
  A \emph{congruence preserving partition} of $\Terms$ is a set $\Partition$ that partitions $\Terms$, such that
  for every $f(\vec{s}), f(\vec{t}) \in \Terms$, whenever $\forall i\in\{1,\dots,n\}.\exists P \in \Partition. s_i, t_i \in P$, then $\exists P' \in \Partition . f(\vec{s}), f(\vec{t}) \in P'$.
\end{Definition}

\begin{Definition}[E-graph, $\Equalities$]\label{def:egraph}
  An E-graph, $\Equalities$, is a data structure that
  provides access to a congruence preserving partition.
  It maps each term $t \in \Terms$ to its partition by $\mathit{root}(t)$.
\end{Definition}

Every interpretation $\Model$ induces a
congruence preserving partition. Conversely, for a set of
equalities over EUF, there is a unique
congruence preserving partition $\Partition$ such that for each
equality $s \simeq t$, $s, t$ belong to the same class in $\Partition$,
and every other partition with that property is obtained by taking unions of classes in $\Partition$.
When a solver constructs an E-graph $\Equalities$, then every equality that is added to $\Equalities$ is justified by a set of literals.
Justification extends to every implied equality and can be represented as a proof tree that uses axioms for EUF.
Thus, when $\mathit{root}(t) = \mathit{root}(t')$,
then the function $\justify{t}{t'}$ returns the set of equality literals
used to establish the equality between $t, t'$ in $\Equalities$.
We use the partition to find a computable term $t' \in T_{\Computable}$ equal to a given term $t$.
To this end, we define a so-called representative function $\reprC$, returning a suitable $t'$ for the given $t$, as well as for the class $t$ belongs to in $\Partition$.

\begin{Definition}[$\reprC(P)$~\cite{EasyChair:10000,DBLP:conf/cav/GarciaContrerasKSG23}]
  \label{def:reprc}
  Given a congruence preserving partition $\Partition$, a \emph{representative}, overloaded as:
  $\reprC : 2^\Terms \rightarrow T_{\Computable} \cup \{ \bot\}$,
  $\reprC : \Terms \rightarrow T_{\Computable} \cup \{ \bot\}$,
  is a function satisfying:
  \begin{enumerate}
  \item $\reprC(t) = \reprC(P)$ for each $P \in \Partition, t \in P$.
  \item $\reprC(P) = \bot$ or
    $\reprC(P) = f(\reprC(\vec{t}_1), \ldots, \reprC(\vec{t}_n)) \in T_{\Computable}$ such that 
    $f(\vec{t}) \in P, \reprC(\vec{t}_i) \neq \bot$ for each $\vec{t}_i$.        
  \item Maximality:
    for every $P\in \Partition$, $\reprC(P) = \bot$ implies that every term $f(\vec{t})$ in $P$ either has $f \not\in\Computable$,
    or $\reprC(\vec{t}_i) = \bot$ for some $i$.
  \end{enumerate}
\end{Definition}

Searching for a realizer of $y$ for $\Literals[y]$ can be performed by
Algorithm~\ref{alg:embpr-euf}.\footnote{We describe the practical optimizations for Algorithm~\ref{alg:embpr-euf}, including finding the ``any'' realizer $\_$, in Section~\ref{sec:implementation}.}

\begin{algorithm}
  Given $\Literals[y]$ over $\Computable\cup\{y\}$, and $\Model$ such that $\Model\models\Literals[y]$,
  return $\langle\EProject, \realizer\rangle$ over $\Computable$ such that either $\EProject \implies \Literals[\realizer]$, or $\realizer=\bot$ and $\EProject\implies\exists y.\Literals[y]$.
  \begin{enumerate}
  \item Let $\Partition$ be a congruence preserving partition, based on $\Model$, satisfying $\Literals$. 
  \item Let $\mathcal{C} := \{ \reprC(P) \mid P \in \Partition, \reprC(P) \neq \bot \}$ \\
    \phantom{Let $\mathcal{C} :=\ $} $\cup\ \{ f(\vec{t}) \mid f\in\Computable, t_i = \reprC(P_i) \neq \bot \}$.
  \item If there is $\realizer\in\mathcal{C}$ such that $\Literals[y] \land y\simeq\realizer$ is satisfiable, return $\langle \Literals[\realizer], \realizer\rangle$.
  \item Otherwise, there is no computable $\realizer$ such that $\Literals[\realizer]$ does not contradict equalities following from $\Model$. Return $\langle \EMBP(\Model, y, \Literals), \bot\rangle$.
  \end{enumerate}
  \caption{$\EMBPR(\Model,y,\Literals)$ for EUF\label{alg:embpr-euf}}
\end{algorithm}

\begin{Proposition}[Correctness and Completeness of Algorithm~\ref{alg:embpr-euf}]\label{prop:embpr-euf}
  Algorithm~\ref{alg:embpr-euf} produces $\EMBPR(\Model,y,\Literals)$ iff one exists, and returns $\langle\EMBP(\Model, y, \Literals), \bot\rangle$ otherwise.
  
\end{Proposition}

\begin{myproof} 
  The goal is to establish if there exists a term $\realizer \in \ComputableTerms$ such that $\Literals[\realizer] \equiv \Literals[y] \land y\simeq\realizer$ is satisfiable.
  For this purpose we enumerate a subset of $\ComputableTerms$ that is necessary and sufficient.
  In particular, it is sufficient to check the set of terms in $\Partition$ that are already represented by computable terms,
  and terms immediately obtained from these terms by using the signature $\Computable$.
  Suppose there is a term $\realizer\in\ComputableTerms$ such that $\Literals[\realizer]$ is satisfiable.
  Then either it is congruent to a term in $\reprC(P), P \in \Partition$, or we can trim it down to a term
  with immediate children in $\reprC(P)$. Suppose $\realizer = f(\vec{t})$ is a solution, but $t_i$ for some $i$ is not congruent
  within $\Partition$. Then we could use $t_i$ in place of $\realizer$ because we can produce a model of $\Literals[t_i]$
  by setting the interpretation of $t_i$ the same as $f(\vec{t})$.
  Note that it is possible that $\Model$ enforces congruences not in $\Partition$ so it is not necessarily the case
  that $\Model \models \Literals[\realizer]$.

\end{myproof}

\begin{Remark}\label{remark:embpr-euf}\ Single invocations of \EMBPR{} do not necessarily produce formulas
  that are equivalent to $\exists y.\Literals[y]$. Thus, $\Literals[y] \land \neg \Literals[\realizer]$ is not necessarily
  unsatisfiable. However, \EMBPR{} for EUF+LIA finitely eliminates all models of $\Literals$.
\end{Remark}

\begin{example}
  In the following constraints, there are no solutions for $y$ using $\ComputableTerms$.
  (1)~Let $\Computable = \{ a\}$ and $\Literals = y \not\simeq a$.
  (2)~Let $\Computable = \{ f, a \}$ and $\Literals = f(a) \simeq a \land y \not \simeq a$.
\end{example}

\begin{example}\label{ex:workshop-embpr}
  We compute $\langle\EProject, \realizer\rangle$ for $\Literals[y] = \{A\simeq y, \Sunday\}$, requested in Example~\ref{ex:workshop}.
  A partition $\Partition$ based on $\Literals$ is $\{\{A, y\}, \{\Sunday, \top\}\}$.
  Then $\mathcal{C} = \{A, \Sunday\}$. %
  Since $A$ is of the suitable sort and $\Literals[A]\land y\simeq A$ is satisfiable,
  the result is $\langle\Sunday\land A\simeq A, A\rangle$, equivalent to $\langle\Sunday, A\rangle$.
\end{example}

Extending MBP from LIA to EUFLIA requires enforcing
Ackermann axioms. We derive a recipe from~\cite{EasyChair:10000}.
\begin{Definition}[\EMBPR{} for LIA + EUF]
  Let $\Model \models \Literals$ over LIA + EUF, and assume $\Literals$
  is closed under Ackermann reductions with respect to $y$.
  That is, for every occurrence $f(s)$ and $f(t)$,
  where $f$ is uninterpreted, $s, t$ are arithmetic terms that contain $y$:
  either $(s \not\simeq t) \in \Literals$ or
  $(f(s) \simeq f(t)) \in \Literals$.
  Then $\EMBPR(\Model, y, \Literals)$ is defined by LIA projection
  where only occurrences of $y$ within arithmetic terms are considered.
\end{Definition}

\begin{Remark}\label{remark:embp} We established complete \EMBP{} and \EMBPR{}
  methods for EUF and EUFLIA. \EMBP{} for combinations of theories that include arrays and algebraic datatypes
  are implemented in Z3.
\end{Remark}

\subsection{Universal Projection}
\label{sec:universal-projection}

Universal projection for EUFLRA was addressed in different guises.
Covering algorithms by~\cite{DBLP:conf/esop/GulwaniM08} can
be used to compute projections for EUFLRA.
They provide a strongest $\Psi$ such that
$(\exists \Uncomputable. \neg \Phi) \implies \Psi$.
Equivalently, $\neg \Psi \implies \forall \Uncomputable. \Phi$.
We would like to unfold $\neg\Phi$ incrementally as a DNF so that we can
compute implicants of $\forall\Uncomputable.\Phi$ incrementally and terminate once
we find a suitable projection. %

We here outline a method that produces a DNF of implicants.
It can be used for ground formulas and combinations of theories where \EMBP{} applies.
It derives from the MARCO algorithm~\cite{DBLP:journals/constraints/LiffitonPMM16} for core enumeration, where our twist is to apply
projection to cores during enumeration and to terminate early when a suitable satisfying assignment is found.
Let $\Phi$ be a ground formula and $\Literals$ be all the literals that occur in the negation normal form of $\neg \Phi$.
The set of cores over $\Literals$ with respect to $\Phi$ can be enumerated using core enumeration techniques.
Each core $\Core$ is such that $\Core \land \Phi$ is unsatisfiable, which means $\Phi \implies \neg \Core$.
We can filter out cores that are unsatisfiable by themselves. 
It remains to eliminate $\Uncomputable$ from each core, or in other words, find $\Core'$,
such that $(\forall \Uncomputable. \neg\Core) \implies \neg\Core'$.
This question is equivalent to finding $\Core'$, such that
$\Core' \implies \exists \Uncomputable . \Core$.
We can use \EMBP{} to establish $\Core'$ based on a model for $\Core$.
Thus,
\[
\begin{array}{llllllll}
  \forall \Uncomputable . \Phi & \equiv & \forall \Uncomputable . (\Phi \land \bigwedge_i \neg \Core_i) \\
  & \equiv & (\forall \Uncomputable . \Phi) \land \bigwedge_i \forall\Uncomputable . \neg \Core_i \\
  & \equiv & (\forall \Uncomputable . \Phi) \land \bigwedge_i \neg \Core'_i 
\end{array}
\]
We have found a sufficient set of cores when (if) the formula 
\[
  (\bigwedge_i \neg \Core'_i) \land \neg \forall \Uncomputable . \Phi
\]
is unsat. But we do not need to compute the full set of cores before we can extract candidates for $\UProject$.
We can return a propositional model $\UProject$ of the current partial
set of projections $\bigwedge_i \neg \Core'_i$, whenever $\UProject \land \neg\forall\Uncomputable.\Phi$
is unsat.
In other words, it suffices to extract a DNF of $\bigwedge_i \neg \Core'_i$
on demand without having the full $\bigwedge_i \neg \Core'_i \implies \forall \Uncomputable.\Phi$.

This discussion motivates Algorithm~\ref{alg:universal-project-loop}.
It enumerates propositional models of literals in $\Phi$, using additional constraints $\ProjectionClauses$ (blocking implicants of $\neg\Phi$) and $\CorrectionSets$ (blocking implicants of $\Phi$ which use uncomputable symbols), both initialized to $\top$.
The models are constructed by first finding a subset $U$ of uncomputable literals in $\neg\Phi$
consistent with $\ProjectionClauses \land \CorrectionSets$,
and then finding a model for $U\land\Phi$.
If this model satisfies a set of literals $\ComputableLiterals^\top$ not using uncomputables,
and $\ComputableLiterals^\top \implies \Phi$, we are done (step~7).
If $U\land\Phi$ is unsat, we block this model in future enumeration:
we project $\Uncomputable$ from the unsat core $\mathit{Core}(U,\Phi)$,
negate the projection obtaining $\bigvee_{\ell\in\bar{\Core}}\ell$ (corresponding to $\neg\Core_i'$ from the discussion above), add this clause to $\ProjectionClauses$,
and add the new literals from $\bar{\Core}$ to all clauses in $\CorrectionSets$ (step 4).
The last case is when the propositional model implies $\Phi$, but using an implicant which includes uncomputable literals.
To block this model in future enumeration iterations,
we add a clause over literals that are false in this model to $\CorrectionSets$ (step 8).
In contrast to MARCO, the set of literals that is used to enumerate cores grows (in step 4).
To allow models using the new literals, the algorithm weakens $\CorrectionSets$ by including new literals from $\bar{\Core}$.
The algorithm is resumable after returning in step 7 by supplying a clause
$\bigvee_{\ell\in\bar{\Core}}\ell$ implied by $\neg\UProject$ to step 4(a).

\

\begin{algorithm}
  Given $\Phi$ and $\Uncomputable$, return $\UProject$ such that either $\UProject\implies\forall\Uncomputable.\Phi$, or $\UProject=\bot$.

  \begin{enumerate}
  \item Let $\ProjectionClauses := \top, \CorrectionSets := \top$,
    let $\UncomputableLiterals$ be literals in $\NNF(\neg \Phi)$ that contain symbols from $\Uncomputable$,
      and let $\ComputableLiterals$ be the literals in $\NNF(\Phi)$ that do not contain symbols from $\Uncomputable$.
  \item If $\ProjectionClauses \land \CorrectionSets$ is unsat, then return $\bot$.
  \item Find a subset $U \subseteq \UncomputableLiterals \cup \ComputableLiterals$ that is satisfiable with model $\Model$
    and implies $\ProjectionClauses \land \CorrectionSets$.
  \item If $U \land \Phi$ is unsat, let $\Core := \mathit{Core}(U, \Phi)$,
    let $\bar{\Core} := \{\bar{\ell} \mid \ell\in \EMBP(\Model, \Uncomputable, \Core)\}$, and
    \begin{enumerate}
      \item
        add $\bigvee_{\ell\in \bar{\Core}}\ell$ to $\ProjectionClauses$,
        update $\CorrectionSets := \{ S \lor \bigvee_{\ell\in(\bar{\Core}\setminus\ComputableLiterals)}\ell \mid S \in \CorrectionSets \}$,
        $\ComputableLiterals := \ComputableLiterals \cup \bar{\Core}$, and
    \item  go to step 2.
    \end{enumerate}
  \item Otherwise, let $\Model'$ be such that $\Model' \models U \land \Phi$.
  \item Let $\ComputableLiterals^\top := \{\ell\in \ComputableLiterals \mid \Model'(\ell) = \top$\}
  \item If $\ComputableLiterals^\top \land \neg \Phi$ is unsat, then $\ComputableLiterals^\top$ is an implicant of $\Phi$. Return $\UProject := \mathit{Core}(\ComputableLiterals^\top, \neg \Phi)$.
  \item Let $\Literals^\bot := \{\ell\in \ComputableLiterals\cup\UncomputableLiterals \mid \Model'(\ell) = \bot\}$, add $\bigvee_{\ell \in \Literals^\bot} \ell$ to $\CorrectionSets$, and go to step 2.
  \end{enumerate}
  \caption{Universal projection using cores, correction sets and \EMBP \label{alg:universal-project-loop}}
\end{algorithm}

\begin{Proposition}[Correctness of Algorithm~\ref{alg:universal-project-loop}]\label{prop:uproject-correct}
  If Algorithm~\ref{alg:universal-project-loop} returns $\UProject\not=\bot$, then $\UProject \implies\forall\Uncomputable.\Phi$.
\end{Proposition}

\paragraph{Of Proposition~\ref{prop:uproject-correct}}
\begin{myproof}
  Since the algorithm can only return $\UProject$ from step 7,
  it means $\UProject = \mathit{Core}(\ComputableLiterals^\top,\neg\Phi)$
  where $\ComputableLiterals^\top$ does not contain symbols from $\Uncomputable$,
  and $\UProject\implies\Phi$, which then entails $\UProject\implies\forall\Uncomputable.\Phi$.
\end{myproof}

\begin{Remark}\label{remark:umbp}
  The algorithm provides a way to incrementally compute cases of a partial function.
  It is complete for theories that admit quantifier elimination.
\end{Remark}

\begin{myproof}
  Claim: the sets $\ComputableLiterals$ obtained from the NNF of $\Phi$
  and $\UncomputableLiterals$ obtained form the NNF of $\neg\Phi$ are sufficient
  for enumerating conflicts with $\Phi$, and $\neg\Phi$, respectively.
  If $L$ is a propositional model such that $L \land \Phi$ is unsat
  then $L' := L \setminus \{ \ell \in L \mid \ell \not\in \UncomputableLiterals, \overline{\ell} \in \UncomputableLiterals \}$
  also satisfies $L' \land \Phi$ is unsat.
  The symmetric claim is:
  If $L$ is a propositional model such that $L \land \neg \Phi$ is unsat, then
  $L' := L \setminus \{ \ell \in L \mid \ell \not\in \ComputableLiterals, \overline{\ell} \in \ComputableLiterals \}$
  also satisfies $L \land \neg \Phi$ is unsat.
  Justification (for the last claim): If $\overline{\ell}$ occurs positive in $\Phi$, but $\ell$ does not occur in $\Phi$ at all, then $\ell$
  does not occur negatively in $\neg \Phi$ and therefore doesn't contribute
  to $L \land \neg \Phi$ being unsat.

  Claim: All propositional models of $\bigwedge_i \neg \Core'_i$ are covered by step 7.
  Here, $\Core'_i$ is a projection
  of $\Core_i$ and $\Core_i$ is a propositional model of $\Phi$.
  It follows as step 4 prunes literal combinations that are propositional models of $\neg\Phi$,
  while step 7 identifies the models for $\bigwedge_i \neg \Core'_i$,
  and step 8 forces enumeration of propositional models to contain literals that belong
  to a correction set of non-models of $\Phi$.  

  Put together, if the underlying theory allows to finitely eliminate the propositional models (in step 4 by quantifier elimination), we eventually reach the model corresponding to the implicant of $\forall\Uncomputable.\Phi$.
\end{myproof}

  \begin{example}\label{ex:fu_is_a}
  Let $\Phi := f(u) \simeq a \implies f(u) \simeq y$ where $\Computable=\{f, a\}$, $\Uncomputable=\{u\}$. %
  Let $U := \{f(u)\simeq a, f(u)\not\simeq y\}$.
  Then $U\land\Phi$ is unsat and
  the core $\Core := f(u) \simeq a \land f(u) \not\simeq y$ with respect to $\Phi$ has the projection $a \not\simeq y$.
  We add $a \simeq y$ to $\ProjectionClauses$, and now $\ProjectionClauses \land \neg \Phi$ is unsatisfiable and we can return
  $a \simeq y$ in step 7.
  Existential projection computes the pair $\langle \top, a \rangle$, which is a solution to the synthesis problem.
\end{example}

\begin{example}\label{ex:workshop-umbp}
  We compute $\UProject$ requested by Algorithm~\ref{alg:skolem-synthesis} in Example~\ref{ex:workshop}.
  We have $\Phi:=\big(
    (\Monday \lor \Sunday) \ \land \
    (\Monday \implies w(\mathit{V})) \ \land \
    (\Sunday \implies w(\mathit{A}))     
    \big) \implies w(y)$
  and $\Uncomputable=\{w\}$.
  First let $U := \{\neg\Monday, \Sunday, w(A), w(V), \neg w(y)\}$.
  Then $U\land\Phi$ is unsat, with $\Core = \mathit{Core}(U,\Phi)=\{\Sunday, w(A), \neg w(y)\}$, projected to $\{\Sunday, A\not\simeq y\}$.
  We let $\ProjectionClauses := \neg\Sunday\lor A\simeq y$ and find a new $U := \{\neg\Monday,\neg\Sunday, w(A), w(V),\neg w(y)\}$.
  Then $U\land\Phi$ is sat, and we find its model $\Model' = \{\Sunday\mapsto\bot, \Monday\mapsto\bot, A\mapsto v_0, V\mapsto v_1, y\mapsto v_2, w\mapsto\lambda v.v\not=v_2\}$.
  We get $\ComputableLiterals^\top = \{\neg\Monday, \neg\Sunday\}$, which implies $\Phi$, and we thus return $\UProject = \mathit{Core}(\ComputableLiterals^\top, \neg\Phi)= \{\neg\Monday, \neg\Sunday\}$.

\end{example}

%% file: unique.tex
\section{Synthesizing Unique Functions}
\label{sec:unique-functions}

When specifications have unique realizers, we can take important shortcuts.
For synthesis tasks based on Definition~\ref{def:partial-synth}, %
there may be unique solutions to $y$ based on implicants $\UProject$ of
$\Phi$. %
In that case, we can bypass invoking $\EMBPR{}$.
It is well-recognized in QBF solving~\cite{DBLP:conf/cav/RabeTRS18}
that there can be a substantial advantage to detecting when a formula entails that there is a unique solution for an existential variable.
Heuristics based on unique solutions are also used to significantly aid synthesis of Boolean functions~\cite{DBLP:journals/amai/AkshayCS24}.
We here outline how we leverage unique function specifications in the case of EUF and LRA.

  The setting we will be discussing is the following: given a variable $y$, we introduce inferences
  that can establish when a set of constraints $\Phi$ entails that $y$ is equal to a term $t$ that does not contain
  any of the symbols from $\{ y \} \cup \Uncomputable$. For LRA in isolation, there
  is a canonical method for inferring equalities from conjunctons of equalities and inequalities:
  Gaussian elimination. For EUF in isolation, we can
  consult $\reprC$ from Definition~\ref{def:reprc} if the congruence closure induced by a conjunction of equalities
  implies that $y$ is congruent to a term without symbols from
  $\{ y \} \cup \Uncomputable$. We summarize these methods in a bit more detail next.
  Our main observation is that the \emph{combinination} of EUF and LRA is also amenable
  to synthesizing unique solutions. In Section~\ref{sec:EUFLRA} we show how to extract
  unique solutions for $y$ (if they exist) in a combination of the two theories.
  The main claim is that Gaussian elimination can be interleaved with congruence closure rules
  to establish all equalities modulo LRA + EUF under a set of asserted literals.
  It builds on an idea of using Gaussian elimination to solve for symbols that
  should be eliminated and use the resulting equations to augment the congruence closure
  for equalities that may have not been surfaced by the LRA solver.
  Recall, that it is enough for theory combinations to agree on equalities with respect to
  a candidate model~\cite{ModelBased} and not all implied equalities. So our method
  extracts implied equalities on the fly from LRA and takes them into account by augmenting the
  congruence closure.
  Beyond experimenting with this method on our benchmarks, this approach also lets Z3 expose
  equality saturation modulo EUF and LRA.

\subsection{EUF}
\label{sec:euf}

The heart of CDCL(T) solvers is a congruence graph $\Equalities$ for the set of terms in $\Phi$.
In a satisfiable state, $\Equalities$ maintains a set of currently implied congruences.
If in a satisfiable state $\reprC(y) \in \ComputableTerms$ and
$\justify{y}{\reprC(y)}$ has no terms from $\Uncomputable$,
then we have a justification that $y$ is equal to a computable term using assumptions that
are computable or use $y$. Assuming the state is consistent for EUF, the justification
$\justify{y}{\reprC(y)}$ is also EUF-consistent.
We then produce the condition $\justify{y}{\reprC(y)}[\reprC(y)/y]$ and the realizer $\reprC(y)$.
By construction, this approach extracts realizers that are forced by $\Equalities$, but it may
fail to detect that a set of constraints have only one possible realizer:

\begin{example}
  Let $\Sigma = \{ a, b \}, \Literals := y \not\simeq b \land a \not\simeq b$. Then $a$ is a unique realizer for $y$,
  but for the congruence class over $\Literals$, $\reprC(y) = \bot$.  
\end{example}

\subsection{LRA}
\label{sec:unique-lra}

It is also possible to solve for variables over LRA.
We recall the dual tableau normal form used by the arithmetic solver.
A tableau $T$ maintains a set of basic,~$\mathcal{B}$, and non-basic, $\mathcal{N}$,
variables such that each basic variable is in a solved form from the non-basic variables.
The sets $\mathcal{B}$ and $\mathcal{N}$ are disjoint.
\[
  \Tableau: \bigwedge_{b \in \mathcal{B}} \Big( x_{b} = \sum_{i \in \mathcal{N}} a_{bi}x_i \Big)
\]
Furthermore, it maintains bounds $\Bounds$ on all variables
$\bigwedge_i \lo{x_i} \leq x_i \leq \hi{x_i}$, where lower $\lo{x_i} \in \{-\infty \} \cup \mathbb{R}$,
$\hi{x_i} \in \{\infty\} \cup \mathbb{R}$.
Every lower and upper bound that is not infinite is justified by a set of literals $\mathcal{J}_{lo}(x)$, $\mathcal{J}_{hi}(x)$.
A feasible tableau comes with an evaluation function $\val{x_i}$ that maps every variable into a value from $\mathbb{R}$, such
that:
\[
\begin{array}{lll}
  \val{x_{b}} = \sum_{i \in \mathcal{N}} a_{bi}\val{x_i} & \mbox{ for every } b \in \mathcal{B}
\end{array}
\]

Let $x$ be an arithmetic variable. It can be reduced to a solution in terms of other variables or constants in the following ways:
\begin{enumerate}
\item $x = \lo{x}$, when $x$ is fixed by existing bounds $\lo{x} = \hi{x}$.
\item $x = \val{x}$, when the current value for $x$ cannot be changed, i.e., when $(x > \val{x} \lor x < \val{x}) \land \Tableau \land \Bounds$ is infeasible.
\item $x = \sum_i a_i x_i$, when $x \in \Tableau$ and $\Tableau', x = \sum_i a_i x_i$ is obtained from $\Tableau$ by pivoting $x$ to a base variable.
\end{enumerate}

\subsection{EUF + LRA}
\label{sec:EUFLRA}

Suppose there is a term $t \in \ComputableTerms$ such that $\Literals$ entails $y \simeq t$.
Based on the ingredients for EUF and LRA we establish a proof search for $t$.
Represent $\Literals$ as $\Equalities + \Tableau + \Bounds$, then the search
succeeds by deriving the sequent $\Equalities, \Tableau, \Bounds \vdash y \leadsto t$ with $t\in \ComputableTerms$, using the following rules:
\begin{gather*}
  \AxiomC{$\Equalities, \Tableau, \Bounds \vdash t_i \leadsto s_i$ for each $i$}
  \UnaryInfC{$\Equalities, \Tableau, \Bounds \vdash \sum_i a_i t_i \leadsto \sum_i a_i s_i$}
  \DisplayProof
\\[1em]
  \AxiomC{$\reprC(t) \neq \bot$}
  \UnaryInfC{$\Equalities, \Tableau, \Bounds\vdash t \leadsto \reprC(t)$}
  \DisplayProof
  \qquad
  \AxiomC{$t$ is fixed by $\Tableau, \Bounds$}
  \UnaryInfC{$\Equalities, \Tableau, \Bounds \vdash t \leadsto \val{t}$}
  \DisplayProof
\\[1em]
  \AxiomC{$\begin{array}{c}
      \Equalities \vdash t = f(\vec{t}), f\in\Computable\\
      \Equalities, \Tableau, \Bounds \vdash t_i \leadsto t_i' \mbox{ for each } i
    \end{array}$}
  \LeftLabel{Congruence}
  \UnaryInfC{$\Equalities, \Tableau, \Bounds \vdash t \leadsto f(\vec{t}')$}
  \DisplayProof
  \\[1em]
  \AxiomC{$\begin{array}{c}
      \Tableau \equiv t = \sum_i a_i t_i, \Tableau'\\
      \Equalities, \Tableau', \Bounds \vdash t \leadsto t'
    \end{array}$}
  \LeftLabel{Gauss}
  \UnaryInfC{$\Equalities, \Tableau, \Bounds \vdash \sum_i a_i t_i \leadsto t'$}
  \DisplayProof
\end{gather*}

\begin{Proposition}\label{prop:euf-lra-unique}
  For a real-valued variable $y$, the search for $t\in\ComputableTerms$ with
  $\Equalities, \Tableau, \Bounds \vdash y \leadsto t$ saturates finitely and
  finds a solution in $\ComputableTerms$ if and only if it exists.
\end{Proposition}

\begin{myproof}[Outline]
  The proof search is finite because every time it solves using $\Tableau$
it removes the row for the solved variable through a Gauss elimination step, 
and the number of terms in $\Equalities$ to check is finite.
Assuming arithmetic operations are computable, the rule that decomposes summations is a special case of the congruence rule.
The proof search is also complete for arithmetical LRA
since if
it is invoked with $y$ and finitely saturates without producing a term for $y$,
we can establish a model where $y$ is distinct from every term in $\ComputableTerms$:
Set $\val{y} := \val{y} \pm \epsilon$ for an infinitesimal $\epsilon$, and inductively, for extensions $t = \Sigma_i a_i t_i$
there is at least one term $t_i$ that cannot be solved for,
so set $\val{t_i} := \val{t_i} \pm \epsilon/a_i$.
The sign $\pm\epsilon$ is chosen to make sure the new value of $y$ is within the bounds for $y$ in $\Bounds$.
Whenever $t_i$ is congruent to $f(\vec{t})$, $f\in\Computable$,
there is inductively an argument $t_i$ which is non-solvable,
so the interpretation of $f$ can be adjusted to take into account the new
value for $t_i$. This produces a non-standard model where the interpretation of all terms in $\ComputableTerms$
are over the rational numbers while $y$ evaluates to a non-standard real.
Functions are interpreted over rationals if all arguments are rationals,
and can otherwise evaluate to a non-standard real if some argument is not a real number.
\end{myproof}

We note that it is straightforward to extend the inference system to also extract justifications
by tracking inequalities used for fixed bounds and equalities used for congruences.

%% file: implementation.tex
\section{Implementation and Evaluation}\label{sec:implementation}
We implemented our method as a proof of concept, and compared it with state-of-the-art methods supporting uncomputables: the synthesis mode of Vampire~\cite{PetraCADE23} and the leading SyGuS solver cvc5~\cite{DBLP:journals/fmsd/ReynoldsKTBD19}, obtaining encouraging early results.

\subsection{Implementation}
We implemented Algorithms~\ref{alg:skolem-synthesis}, \ref{alg:embpr-euf}, and~\ref{alg:universal-project-loop}, as well as the unique solution finding of Section~\ref{sec:unique-functions}, as a stand-alone Python prototype (available online at~\cite{synthesiz3doc}) using the Z3 API, which we also extended by new features.
We use variables of array sort for higher-order quantification.
The theory of arrays in Z3 coincides with SMT-LIB~2~\cite{SMTLIB} for ground formulas,
but not quantified formulas.
E.g., the first-order theory of array select function admits models where select does not include bijections,
but with Z3 one can instantiate quantified array variables by $\lambda x. x$ and select is just a function application.

The API already provided functions for computing \EMBP{}, an incomplete method for \EMBPR,
and accessing the E-graph. %
We extended Z3 by exposing E-graph justifications, and by a new heuristic for
finding witnesses from $\ComputableTerms$ for EUF in \EMBPR. It searches for an E-match for $y$
that does not lead to equating shared sub-terms, and specifically avoids merging sub-terms used in disequalities.
We also added a method for solving variables that is complete for LRA and extends to LIA (incompletely, not to deal with non-convexity of LIA).

We optimize all our algorithms as follows.
In Algorithm~\ref{alg:skolem-synthesis}, step 6, if $\Uncomputable=\emptyset$, we use $\Phi[\realizer/y]$ as the condition for realizer $\realizer$ instead of $\EProject$.
Further, to compute $\EMBPR(\Model, y, \Literals)$ of Algorithm~\ref{alg:embpr-euf}, we first check
if $\Literals$ contains $y$, and if it does not, we return the ``any'' realizer (as in Example~\ref{ex:workshop}).
Otherwise, we try calling \EMBPR{} of the API, and
if that does not return a realizer, we iterate over the candidates provided by $\reprC$, only producing a new one if the last did not work out.
Finally, in Algorithm~\ref{alg:universal-project-loop}, step 7, we check the unsatisfiability with respect to $\lnot \Phi'$, where $\Phi'$ is (i) initialized to the original specification $\Phi$, and (ii) only if Algorithm~\ref{alg:universal-project-loop} returns the same $\UProject$ twice, updated to $\Phi' := \Phi\land\lnot C$ using the current $C$ of Algorithm~\ref{alg:skolem-synthesis}.

\subsection{Evaluation}
\begin{table}
  \newcolumntype{C}[1]{>{\centering\let\newline\\\arraybackslash\hspace{0pt}}m{#1}}
  \renewcommand{\arraystretch}{1.3}
  \setlength{\tabcolsep}{2pt}
  \begin{tabular}{cp{4.5cm}cC{0.8cm}C{0.8cm}C{0.95cm}}
    Id & Benchmark subset description & Count & Has $\Uncomputable$ & Partial & Unique \\ \hline \noalign{\vskip 0.2em} 
    \sygusbench & simple if-then-else benchmarks from SyGuS competition~\cite{sygus-comp} & 3 & \xmark & \xmark & \cmark \\
    \knapsackbench & $k$-knapsack (see Example~\ref{ex:running1}) & 5 & \cmark & \xmark & \cmark \\
    \maxbench & maximum function for $n$ given integers & 10 & \xmark & \xmark & \cmark \\
    \lowernsbench & lower bound function for $n$ integers & 10 & \xmark & \xmark & \xmark \\
    \lowersbench & function finding an integer lower than all $n$ given integers & 10  & \xmark & \xmark & \xmark \\
    \betweentbench & function finding an integer between the given bounds, \underline{assuming} some exists & 4 & \cmark & \xmark & \xmark \\
    \betweenpbench & function finding an integer between the given bounds, \underline{if} some exists & 4 & \cmark & \cmark & \xmark \\
    \equationtbench & fuction finding a solution for $y$ in the given equation, \underline{assuming} it exists & 3 & \cmark & \xmark & \cmark \\ 
    \equationpbench & function finding a solution for $y$ in the given equation, \underline{if} it exists & 3 & \xmark & \cmark & \cmark \\ 
    \uftbench & Examples~\ref{ex:workshop-spec} (and a computable variant thereof) and \ref{ex:fu_is_a} (and a variant thereof)  & 4 & \xmark+\cmark & \xmark & \cmark \\ %
    \ufpbench & the first specification or Example~\ref{ex:qy_or_aby} & 1 & \cmark & \cmark & \cmark  %
  \end{tabular}
  \caption{Summary of the 57 benchmarks used: set id, description, how many benchmarks are in the subset, and whether they use uncomputables, specify a partial function, and a unique function.
  Benchmark sets \sygusbench{}-\equationpbench{} use LIA, \uftbench{} and \ufpbench{} use UF.
  In set~\knapsackbench{} we have $k\in\{1,2,3,4,5\}$, and in sets \maxbench-\lowersbench{} we have $n\in \{2, 5, 10, 15, 20, 25, 30, 50, 100, 150\}$.}\label{tab:benchmarks}
\end{table}

To the best of our knowledge, there was no standard benchmark set for synthesis with uncomputable symbols as of writing this paper.
However, a dataset based on the benchmarks from~\cite{PetraCADE23} was under development in parallel with this paper, and in the meantime it was accepted for publication~\cite{CICM25}.\footnote{The dataset was co-developed by one of the authors of this paper.
It is available at \url{https://github.com/vprover/vampire_benchmarks/tree/master/synthesis}.
The version that was available shortly before this paper was submitted, and which we used for evaluation, is at \url{https://github.com/vprover/vampire_benchmarks/tree/6356e52/synthesis}.}
We took the subset of this dataset supported by our
method,\footnote{We excluded problems using (i) theories beyond EUFLIRA, or
(ii) more complex quantification, or
(iii) interpreted uncomputables, or
(iv) which were too similar to others in the dataset -- e.g. we used maximum but not minimum.}
and extended it with the solvable examples from this paper, as well as additional %
partial function synthesis problems. %
Table~\ref{tab:benchmarks} shows a summary of the benchmarks, organized into 11 subsets denoted I-XI.
The set contains 57 benchmarks: 18 small, conceptually distinct problems (all 3 problems of subset~\sygusbench, 2 per both~\betweentbench{} and~\betweenpbench, 4 in~\uftbench, and one per all the other subsets), and 39 scaled-up versions of the small instances.

\begin{table}
  \newcolumntype{C}[1]{>{\centering\let\newline\\\arraybackslash\hspace{0pt}}m{#1}}
  \setlength{\tabcolsep}{2pt}
  \begin{tabular}{lcccccccccccr}
    & \sygusbench & \knapsackbench & \maxbench & \lowernsbench & \lowersbench & \betweentbench & \betweenpbench & \equationtbench & \equationpbench & \uftbench & \ufpbench & \ \ \ sum total \& partial \\ \hline \noalign{\vskip 0.2em}
    Synthesiz3 \ \ & \bf 3 & \bf 3 & \bf 9 & 4 & 3     & \bf 4 & \bf 4 & \bf 3 & \bf 3 & \bf 4 & \bf 1 & \textbf{33} \& \textbf{8} \phantom{rtial}\hspace*{0.11em}\\
    Vampire        & \bf 3 & 0	   & 5	   & 2	   & 0	   & \bf 4 & --    & 0	   & -- 	 & 2	   & --	   & \ \ 16 \& --  \phantom{rtial}\hspace*{0.11em}\\
    cvc5           & \bf 3 & 1     & 5     & \bf 5 & \bf 5 & \bf 4 & --    & 0     & --    & \bf 4 & --    &  27 \& -- \phantom{rtial}\hspace*{0.11em}
  \end{tabular}
  \caption{Experimental results: numbers of problems solved by each solver in benchmark sets I-XI.
  Benchmark sets \betweenpbench, \equationpbench, and \ufpbench{} specify partial synthesis problems not supported by Vampire nor cvc5, which we denote by ``--''.}\label{tab:results-horizontal}
\end{table}

We evaluated the performance of our prototype on this dataset and for the total synthesis problems also compared it to Vampire and cvc5.\footnote{We used Vampire v4.9, commit \texttt{dde88d57f}, in the configuration \texttt{-{}-decode} \texttt{dis+32\_1:1\_tgt=off:qa=synthesis:ep=off:alasc\linebreak a=off:drc=ordering:bd=off:nm=0:sos=on:ss=included:si\linebreak =on:rawr=on:rtra=on:proof=off:msp=off\_600}, and cvc5 v1.1.2 in the default configuration.
For cvc5, we used SyGuS translations of the problems, encoding uncomputables as non-input variables.}
We ran each benchmark on a single core of a 1.60GHz CPU with 16 GB RAM, with the time limit of 1 minute.
We chose 1 minute as it is enough time to solve the solvable small problems, and show the trend for the scaled-up problem instances, but not solve all the scaled-up instances.
Table~\ref{tab:results-horizontal} shows the numbers of solved problems. %
Successful solving of the small instances took all solvers less than 1 second.
For the larger instances, the runtimes increased up to the time limit.
We note that what makes many scaled-up instances hard is that they encode problems with many symmetries: e.g., each case of the maximum-of-$n$-integers function has the same structure, but with different variables.

Overall, our approach solves the most total function synthesis problems, and also partial synthesis problems which cvc5 and Vampire do not support.
Our approach scales as well as or better than both Vampire and cvc5 for all benchmarks with a unique solution, and slightly worse than cvc5 on problems with infinitely many possible solutions (sets~\lowernsbench{} and~\lowersbench).
Notably, within 1 minute, only our approach solves $2$- and $3$-knapsack, and it also finds the maximum-of-$n$-integers function for $n$ up to $100$, while cvc5 and Vampire can go only up to $n\leq 20$.
Interestingly, both cvc5 and Vampire fail to solve all benchmarks of set~\equationtbench, including the smallest instance $\langle 2u\simeq x \implies 2y\simeq x, \{u\}, y\rangle$, ``find the half of an even $x$''.

%% file: related.tex
\section{Related Work}\label{sec:related}
\paragraph*{Synthesis}
The field of synthesis encompasses very different approaches.
We only overview methods similar to ours: synthesizing functions from scratch based on specifications given as logical formulas with some syntactic restrictions.
The general restriction setting in~\cite{DBLP:journals/logcom/Tammet95}
is to integrate a black-box function within a solver that determines whether a term is computable.
We base our specification format on~\cite{PetraCADE23}.
In contrast with~\cite{PetraCADE23}, we distinguish uncomputable symbols directly in the specification formula by using quantification.
The deductive synthesis methods of~\cite{PetraCADE23,DBLP:journals/logcom/Tammet95} are based on theorem proving, and~\cite{PetraCADE23} is implemented in Vampire~\cite{Vampire13}.
While these methods introduce specialized inference rules to filter terms using uncomputable symbols, our approach uses quantifier elimination games to remove uncomputables, and thus requires no changes in the underlying search procedures.

Another related paradigm is SyGuS~\cite{DBLP:series/natosec/AlurBDF0JKMMRSSSSTU15}, which %
supports specifying multiple functions with different inputs, and constraining the space of solutions to functions generated by a given grammar.
SyGuS solvers include cvc5~\cite{cvc5,DBLP:journals/fmsd/ReynoldsKTBD19} and DryadSynth~\cite{dryadsynth}.
Compared to the specification setting we consider, SyGuS allows a more fine-grained syntactic control over the synthesized function. %
We note that to solve such specifications, we could generate candidate terms using the provided grammar instead of using \EMBPR, or that we could use grammar-based reconstruction as in~\cite{DBLP:journals/fmsd/ReynoldsKTBD19} on top of \EMBPR.
On the other hand, our specifications support the use of uninterpreted symbols, %
in particular uninterpreted functions and predicates.
This allows us to express e.g. the workshop problem (see Example~\ref{ex:workshop-spec}), which is not directly supported by SyGuS.
However, using higher-order logic, uninterpreted functions can be encoded as higher-order variables, and with this encoding cvc5 solves the workshop problem (and some other problems, see Section~\ref{sec:implementation}).
Finally, neither SyGuS nor the deductive synthesis methods support partial function synthesis.

We provide some examples where the partial function synthesis problem has no solutions.
The study of unrealizability~\cite{DBLP:journals/pacmpl/NagyKRD24}
dually establishes methods to show that there are no solutions to synthesis problems.

\paragraph*{Projections}
The literature on interpolation, including~\cite{DBLP:conf/aplas/GuptaPR11},
develops algorithms based on proof objects.
Symbol elimination in saturation~\cite{DBLP:conf/cade/KovacsV09} works by
deriving clauses that are consequences of
$\exists \Uncomputable. \neg \Phi$.
Negations of these clauses can be used as projections.
The approach of~\cite{DBLP:conf/cade/KovacsV09} relies on symbol orderings to eliminate symbols through
rewriting, but does not offer guarantees of complete symbol elimination.
Deductive synthesis with uncomputables~\cite{PetraCADE23} blocks inferences which would lead to uncomputable realizers,
but also is not complete.
As an example, inferences such as
$f(y) \not\simeq f(a) \vdash y \not\simeq a$ with $\Uncomputable = \{f\}$
are not part of this methodology.\footnote{We note that such an inference could be supported
by a suitable variation of unification with abstraction~\cite{UWA-Vampire}.}

Notably, MCSAT~\cite{DBLP:conf/cade/JovanovicM12} relies on a dual to $\EMBP$,
called $\UMBP$, in a setting where it is used for satisfiability checking of quantifier-free formulas.
The starting point for $\UMBP$ is a partial model of a subset of symbols that cannot be completed to a full model
of all symbols. A combination of \UMBP{} and \EMBP{} for quantifier solving for polynomial arithmetic is developed
in~\cite{DBLP:journals/corr/abs-2411-03070}, while \EMBP{} with cores
and strategies are used in~\cite{DBLP:conf/lpar/BjornerJ15,DBLP:conf/cade/BonacinaGV23,DBLP:conf/cav/MurphyK24}

We note that stand-alone projection is not necessarily the most
efficient method for solving quantifiers. Templates have recently been used successfully for
non-linear arithmetic~\cite{QuantifierTemplate}, and Z3's Horn clause solver uses
information from several projection invocations to compute more general
projections~\cite{DBLP:journals/fmsd/KrishnanCSG24}.

\paragraph*{Implied equalities}
Bromberger and Weidenbach~\cite{BrombergerW16} present 
a method for extracting all implied equalities from a set of linear inequalities over the reals.
It is used when solving over LIA.
Z3~\cite{DBLP:conf/cav/BjornerN24} uses an incomplete heuristic to detect implied equalities
during search. The QEL algorithm~\cite{DBLP:conf/cav/GarciaContrerasKSG23}
uses equalities that are consistent with an interpretation, but not necessarily implied by a current assignment to literals,
to simplify quantifier elimination problems.